# Temporal convolutional neural networks to generate a head-related impulse response from one direction to another

Tatsuki Kobayashi[1], Yoshiko Maruyama[1,2], Isao Nambu[1], Shohei Yano[3], Yasuhiro Wada[1]

1 Graduate School of Engineering, Nagaoka University of Technology, Japan
2 National Institute of Technology, Hakodate College, Japan
3 National Institute of Technology, Nagaoka College, Japan

**Abstract**

Virtual sound synthesis is a technology that allows users to perceive spatial sound through headphones or earphones. However, accurate virtual sound requires an individual head-related transfer function (HRTF), which can be difficult to measure due to the need for a specialized environment. In this study, we proposed a method to generate HRTFs from one direction to the other. To this end, we used temporal convolutional neural networks (TCNs) to generate head-related impulse responses (HRIRs). To train the TCNs, publicly available datasets in the horizontal plane were used. Using the trained networks, we successfully generated HRIRs for directions other than the front direction in the dataset. We found that the proposed method successfully generated HRIRs for publicly available datasets. To test the generalization of the method, we measured the HRIRs of a new dataset and tested whether the trained networks could be used for this new dataset. Although the similarity evaluated by spectral distortion was slightly degraded, behavioral experiments with human participants showed that the generated HRIRs were equivalent to the measured ones. These results suggest that the proposed TCNs can be used to generate personalized HRIRs from one direction to another, which could contribute to the personalization of virtual sound.

# 1. Introduction

In our daily life, humans are able to perceive sound in three dimensions by using our left and right ears to judge the direction and distance of the sound sources. This is made possible by the fact that spatial transfer characteristics of sound, such as arrival time, sound levels, and frequency spectrum, differ at each ear. One aspect of these characteristics is known as head-related transfer functions (HRTFs) [1-3]. By convolving the sound sources with HRTFs, spatial virtual sound can be generated through earphones or headphones, as if it were coming loudspeakers. This technique has been used in studies of a virtual auditory display [1,3] and has also been applied to the rehabilitation of hearing loss [4]. It is also expected to be used in entertainment, with several commercially available systems for audio systems and games [5,6].

Accurate virtual sound systems face two challenges [2,7-9]. First, the HRTF is influenced by the shapes of the head, pinna, and upper body, and therefore it is necessary to obtain HRTFs appropriate for each individual (user-specificity). Ideally, HRTF measurement in a special environment is required [10], but to reduce the burden of measurement, previous studies have proposed several solutions. One solution is to use anthropometric features [11-13], which depend on the user's ear, head, and torso and can be used to estimate the best-fitted HRTF without measurement. Another possible solution is to select HRTFs from existing data [8,14,15]. For example, one study proposed selecting HRTFs suitable for the user from a database based on the user's evaluation [14]. Another study used reinforcement learning and user evaluation to improve HRTF [16]. However, the performance of the virtual sound is not guaranteed due to the difference in HRTFs between the database and individuals, or the increased training time required to find the appropriate HRTFs using the user evaluation method.
Another challenge in achieving accurate virtual sound systems is the direction-dependent property of the HRTFs. Since the HRTFs vary for each direction, higher resolution virtual sounds require a larger number of HRTFs. Even with the previously mentioned methods using databases or user evaluation, it can take a long time to obtain appropriate HRTFs, leading to user fatigue and head or pinna misalignment. Therefore, there is a need for a method that can determine user-specific HRTFs with minimal burden.

In this study, we proposed a novel method to generate HRTFs from one direction to another using temporal convolutional neural networks (TCN) [17]. The proposed method is aimed at reducing the effort of HRTF measurements and avoiding long calibration times when preparing multiple directions of virtual sound. Recently, researchers have explored the use of neural networks (including deep neural networks) for HRTF generation. Several studies have used simple neural networks [18,19], auto-encoders [20-22], or convolutional neural networks (CNN) [23-25]. In this study, we adopted TCNs for generating HRTFs because it extracts features embedded in temporal information [17]. In addition, unlike other studies using deep neural networks, we proposed a method focused on HRTF generation from measured HRTFs from one direction. This approach assumes that the measured HRTF or well-localized non-individualized HRTF has already been obtained for one direction. The TCNs are used to generate HRTFs from the other directions where no measurement has been conducted. The effectiveness of the method was evaluated by comparing the measured and generated HRTFs and conducting experiments with human participants to evaluate localization accuracy.

## 2. Material and methods

The proposed method uses neural networks to generate HRTFs for directions that have not been measured. The general procedure involves two steps: (1) training the neural network to learn the relationships between a pair of input HRTF (0 degree) and a HRTF on target direction (Fig. 1a), and (2) using the trained network to generate new HRTFs from the one that has been measured (Fig. 1b and 1c).

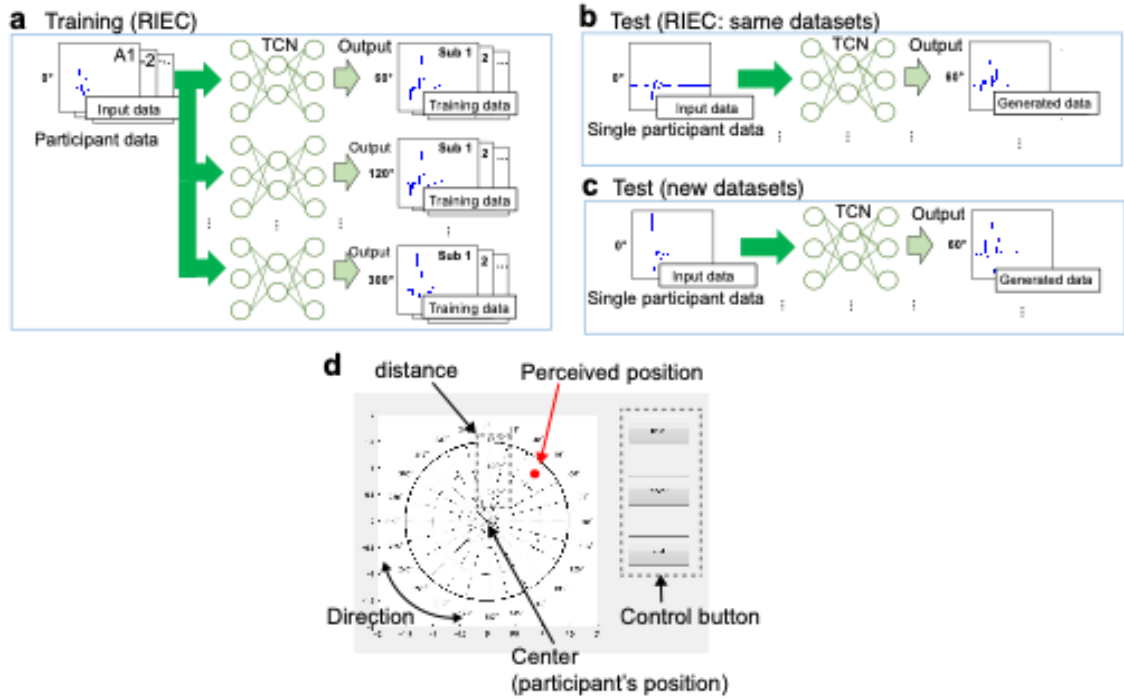

Fig. 1. The proposed method and experimental setting. (a) training of the proposed neural networks, (b) testing on RIEC dataset (within the same dataset of training), and (c) testing on newly measured dataset. (d) Display of the localization test.

### 2.1 HRTF dataset (RIEC)

We used the publicly available RIEC database (Tohoku University, Japan [10]) to train our neural network models. The RIEC data included HRIRs for both ears of 105 participants, with each participant having HRIRs at interval of 5° for azimuth and elevation. The dataset consisted of 512 samples with a sampling frequency of 48 kHz. As a preprocessing step, we down-sampled the data to 44.1 kHz, resampled to 492 samples, and bandpass filtered between 0.2 kHz and 14 kHz to match the conditions of the newly measured data (see below). We only used the data for elevation angles of 0° and azimuth directions that were 60° apart (0°, 60°, 120°, 180°, 240°, and 300°). We only use this experimental setting because our purpose is to generate HRTFs for different targets. The resulting dataset was of size 2 (left and right) × 6 (directions) × 492 (samples) × 105 (participant).

**2.2 HRTF measurement (new dataset, NUT)**

To check the generalization of the proposed method to a new dataset, we conducted an experiment with human participants and measured their HRTFs. The measurement (i.e., ten participants) was carried out in a sound-proof room at the Sound and Vibration Engineering Center of Nagaoka University of Technology. The loudspeakers (SD-0.6, Soundevice, Japan) were positioned 1.5 m away from the participant and set for 24 directions with intervals of 15°. A miniature microphone (XE-1F, Rion, Kokubunji, Tokyo, Japan) was placed at the entrance to the participants' ear canals to measure white noise (100 Hz-15 kHz) with a sound pressure level of 65 dB. The sampling frequency was 44.1 kHz, and the data were bandpass filtered from 0.2 kHz to 14 kHz. Using this experimental setup, we measured the ear-canal transfer function and sound spatial functions [9,26]. Note that for the experiment and analysis, we used HRTFs only for six directions (0°, 60°, 120°, 180°, 240°, and 300°), and our HRTF definition is different from that used in other works (e.g., [2]).

**2.3 Neural network architectures**

For this study, we utilized temporal convolutional neural networks (TCN) [17], a variant of convolutional neural networks (CNN) specifically designed to handle time-series data. TCN has demonstrated quick learning and high accuracy compared to recurrent neural networks.

CNNs typically consist of input, convolutional, pooling, and fully connected layers. Convolutional layers and pooling layers can extract useful information for classification from 2-dimensional data such as images. However, TCN is designed to handle time-series data by incorporating residual blocks, dilated causal convolutions, and dropout. Dilated causal convolution is a combination of causal convolution and dilated convolution, where the former considers the data corresponding to the past in the time-series, and the latter performs convolution with inputs that are spatially spaced.

The parameters for this architecture are listed in the Table 1, and were set based on a previous study [17]. We used the Adam optimizer [27] and searched for the optimal values of the number of channels, layers, and weight decay using the Optuna library [28].

**Table 1.** Parameters of architecture

| Parameter | Value |
|---|---|
| Kernel size | 2 |
| Number of channels | 10–150 |
| Number of layers | 2–8 |
| Weight decay | 1e-6 – 1e-4 |
| Activation function | SiLU, identity function |
| Optimization algorithm | Adam |
| Dropout ratio | 0.2 |
| Number of epochs | 10000 |

SiLU: sigmoid-weighted liner unit. SD: spectrum distortion. SDR: signal to deviation ratio.

We used a combination of spectrum distortion (SD) [13] and signal to deviation ratio (SDR) [29] as the cost function. SD is a metric used to measure the difference between two HRTFs in frequency domain, calculated as follows:

$$SD[\text{dB}] = \sqrt{\frac{1}{M}\sum_{m=1}^{M}\left(20\log\frac{|H[fm]|}{|\hat{H}[fm]|}\right)^2}$$

where M is the maximum frequency bin, *fm* is the frequency at m-th bin, and $H$ is the amplitude of measured HRTF, and $\hat{H}$ is the amplitude of the generated HRTF. The range of frequency for calculating SD was set to 0.2 kHz to 14 kHz, corresponding to the audible range for humans [30].

On the other hand, SDR calculates the difference between two HRTFs in time domain, as follows:

$$SDR[\text{dB}] = 10\log\frac{\sum_{n=1}^{N}h^2[n]}{\sum_{n=1}^{N}(h[n]-\hat{h}[n])^2}$$

where *N* is the length of HRIR, $h$ is the sample of HRIR and $\hat{h}$ is the sample of generated HRIR.

The generated HRTF is closer to the measured value when SD is small, while a large SDR indicates that the two HRIRs are similar. To combine these two different metrics, we selected the cost functions as follows:

$$e = \sqrt{\frac{1}{M}\sum_{m=1}^{M}\left(20\log\frac{|H[fm]|}{|\hat{H}[fm]|}\right)^2} + 10\log\frac{\sum_{n=1}^{N}(h[n]-\hat{h}[n])^2}{\sum_{n=1}^{N}h^2[n]}$$

In this cost function, the first term represents SD while the second term is inversely related to SDR. The number of epochs was set to 10000 to ensure convergence of the cost function.

*2.3.1 The parameter optimization*

We evaluated the generated HRTR in following three ways: (1) Evaluation within the same dataset, (2) Evaluation to new dataset, and (3) Localization experiment with human participants.

*2.3.2 Evaluation within dataset (RIEC)*

First, we conducted an analysis within the RIEC database using 4-fold cross-validation. In each fold-set, the RIEC data were divided into 84 training and 21 test data. The training data were used for training the neural networks and selecting their parameters. The test data were used for generating the HRTF. We repeated these analyses by changing the training and test data for each fold as part of the cross-validation process. Within the training data, we further divided the data into 63 training and 21 validation data and calculated the cost function for each parameter set as an inner-fold cross-validation (4-fold). The validation data were used for evaluating the parameters, and we selected the parameter set that showed the lowest cost function as the optimal parameters.

*2.3.3 Evaluation to new dataset (NUT)*

To evaluate the generalization of the learned networks, we tested whether they could be applied to new dataset that differ from the original RIEC database. We measured the HRTF of the new dataset and compared it with the generated HRTFs using the proposed method. In this evaluation, we used all the data from the RIEC database for training, employing a 5-fold cross-validation where 84 out of 105 data were training data and the remaining 21 data were used as validation data to find hyperparameters. Note that the number of test data (21) are the same, but number of folds was different compared with evaluation in RIEC. This cross-validation was repeated 50 times, and the parameters yielding the lowest cost function were selected as the best parameters for each direction separately. After finding the best parameters, we used the measured data as inputs of the test data.

*2.3.4 Localization experiment with human participants*

Finally, we evaluated the generated HRTFs in a human experiment. For this experiment, we compared the generated HRTFs for evaluation to new dataset with the individual HRTFs measured for the participant. We asked the participant to evaluate the direction and distance of perceived sound using a graphical user interface, as shown in Fig. 1d. The sound was generated using either the generated or measured HRTF, with the order and direction of the sound pseudorandomized. The participant was not informed about the HRTF type (generated or measured) during the experiment.

Ten human participants (s1 to s10) in the newly measured dataset also participated in this experiment after being given instructions and an explanation of the experiment and signing their informed consent. The experiment was approved by the ethical committee of Nagaoka University of Technology according to the guideline of Declaration of Helsinki (No. R3-12). A total of 10 trials for each direction were conducted.

We evaluated the localization error and front-back confusion ratio. The localization error was measured as the absolute difference between the target direction and the perceived direction of the virtual sound. The front-back confusion ratio was calculated as the ratio of the number of front-back confusions to the total number of trials, and then multiplied by 100 to obtain a percentage value.

## 3. Results

To evaluate the effectiveness of the proposed method, we generated HRTFs for five different directions from a single direction. Specifically, we used the HRTF data for the 0-degree direction as input and generated HRTFs for five different directions (60°, 120°, 180°, 240°, and 300°).

### 3.1 Evaluation of data in the dataset

We first evaluated the generated HRTF using data from the RIEC database, which was also the dataset used for training. We examined the performance of the cost function with respect to both SD and SDR metrics separately.

### 3.2 HRTF differences (SD)

Figure 2a shows the results of SD analysis for each direction, where the input was 0° data. Each result was obtained by averaging across participants. Initially, the SD for the training data was more than 20 dB, and the average SD across participants was 40.7 dB. However, it gradually decreased to less than 10 dB by 1,000 epochs, and the results for all directions showed a similar decrease. The SD results for the test data of the same dataset were similar to those of the training data (Fig. 2, middle). The average SD across participants was 5.2 dB for the training data and 6.4 dB for the test data of the same dataset at the end of 10,000 epochs. Thus, the SD for the test data reached that of the training data.

Next, we applied the newly measured HRTF to this neural network. As shown in the right panel of Fig. 2a (test data), the SD showed a decrease as the epochs progressed. The initial average SD was 42.3 dB, but it gradually decreased to 12.6 dB at the final epoch. Although this SD is higher than that obtained from the test data of the same dataset (Fig. 2a, middle), these results indicate that the generated HRTFs are getting closer to the individual ones in terms of frequency spectrum.

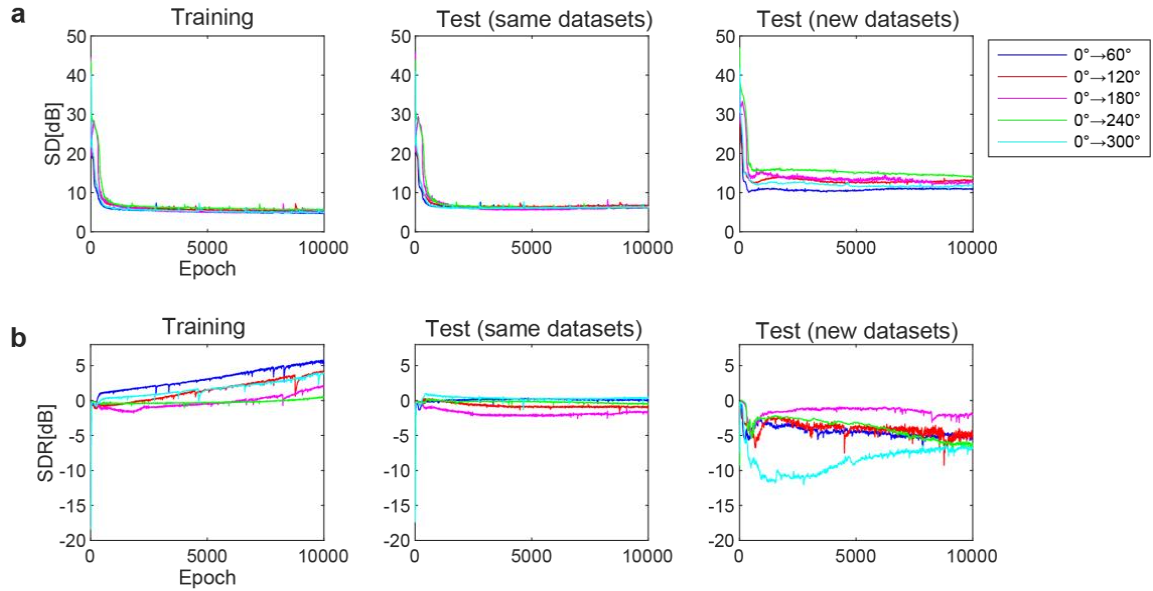

Fig. 2. (a) Average SD, and (b) average SDR across participants and ears. Horizontal axis indicates training epochs and vertical axis denotes SD or SDR. Left panels show SDs and SDRs for training data. Middle panel shows SDs and SDRs for test data within the same dataset (RIEC). Right panel shows SDs and SDRs for test data of new dataset (NUT). Each color represents results for each direction. Input was 0° for all the directions.

### 3.3 HRIR differences (SDR)

In the time domain, we examined SDR for generated HRIRs in each direction (Fig. 2b). When the learning started, SDR was -14.3 dB for training and -13.6 dB for test within the same dataset. With learning progressed, SDR increased and reached an average of 3.2 dB across participants. SDR for the test data (same dataset) also increased to an average of -0.5 dB across participants. This indicates that the generated HRIRs closer to the measured individual HRIRs in the time domain.

For the test data outside of the training data, the initial SDR was -5.4 dB and the final SDR averaged across participants was -5.1 dB. However, the SDR for new dataset did not increase and rather decreased slightly.

### 3.4 HRIR and HRTF examples

Next, we analysed the temporal profiles of HRIRs for the training, test (same dataset), and test (new dataset). Figure 3 displays an example of input (0°, green), measured (60°, blue), and generated (60°, red) HRIRs for each ear. The target direction (ground truth) was on the right side (60°). For the training and test (same dataset) of the left ear (Fig. 3a, left panel), the generated and measured HRIRs are similar, indicating that the proposed method effectively generated HRIRs from 0° data. For the test (new dataset), although the input HRIR changed to match the ground truth, the amplitude of the generated HRIR was lower. For the left ear, the generated HRIR became similar to the measured HRIR, but the amplitude or onset of the impulse is not replicated (Fig. 3b). The results show that the proposed method achieved a good prediction of HRIR for the test data from the same dataset as the training data. However, for the test data from new dataset, there was a tendency of small amplitudes in the generated HRIRs.

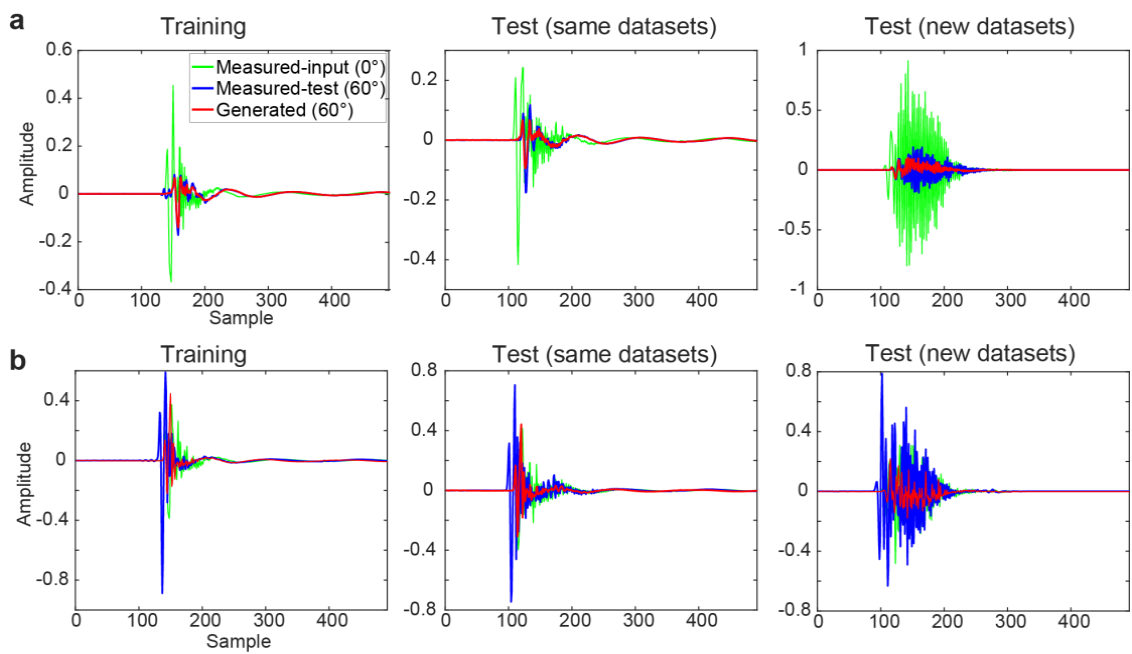

Fig.3 An example of generated HRIRs for representative participant for: (a) left ear and (b) right ear. Left panel shows HRIR for the training dataset. Middle panel shows HRIRs for test participants in the dataset used for training. Right panels show HRIRs for the test participant (s1) in the new dataset. Target angle (direction of the test data) was 60°.

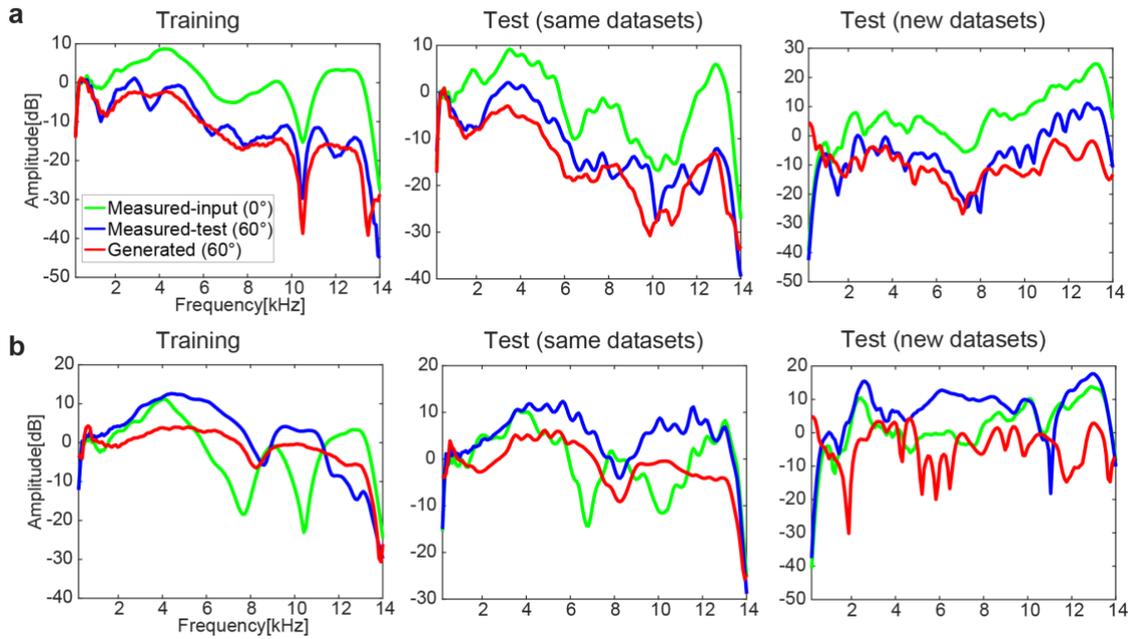

Fig.4 An example of generated HRTF for representative participants for: (a) left ear and (b) right ear. Left panel shows HRTF for the representative participant in the training dataset. Middle panel shows HRIRs for test participant in the dataset used for training. Right panels show HRIRs for the test participant in the new dataset (s1). The same participants' examples as Fig. 3 are shown.

The HRTF profiles generated by our proposed method were found to be close to the measured HRTFs (Fig. 4). For the left ear (Fig. 4a), even when the input HRTFs (green lines in Fig. 4) were different from the target HRTF (measured-test HRTF at 60°, blue lines in Fig. 4) for each participant, the generated HRTFs (red lines in Fig. 4) were apparently close to the target HRTF. For the right ear (Fig. 4b), the generated HRTFs were found to be closer to the target HRTFs than the input, and the notch around 8 Hz was replicated (Fig. 4b left and middle). However, there were still some differences between the target and generated HRTFs even for the training data (Fig. 4b left). Moreover, large errors were observed for some participants in the new dataset (Fig. 4b right).

**3.5 Localization test**

The results mentioned above showed that the proposed method successfully generated HRTFs (HRIRs) that were similar to the ground truth (measured HRTFs). However, it remained unclear whether these generated HRTFs were sufficient for accurate virtual sound localization. To assess this, we conducted an experiment with human participants and evaluated both the localization accuracy and the front-back confusion ratio for the generated HRTFs.

Figure 5 shows the localization errors for each direction and the average across directions. The localization errors for the generated HRTFs increased at 120° (Fig. 5f), 240° (Fig. 5d), and 300° (Fig.5b), while it decreased at 60° (Fig. 5c) and 180° (Fig. 5E), compared to the measured individual HRTFs (ground truth). On average, the errors for the generated HRTFs and the ground truth were similar (Fig. 5A). A statistical test using two-way ANOVA with factors of HRTF type and direction revealed that errors were significantly different for the factor of sound directions ($p < 0.05$) and the interaction of HRTF types and directions ($p < 0.001$). However, no significant difference was observed for the HRTF types ($p > 0.9$), indicating that the generated HRTFs did not degrade localization accuracy.

We also examined the front-back confusion ratio (Fig. 6). The generated HRTFs caused an increase in front-back confusion for some directions and some participants. On average across directions, many participants showed an increase in the front-back confusion ratio. However, no significant difference was found (two-way ANOVA, $p > 0.3$) between generated HRTFs and ground truth, while a difference was observed between directions ($p < 0.05$) and the interaction between HRTF types and directions ($p < 0.001$). This tendency is similar to the results of localization errors.

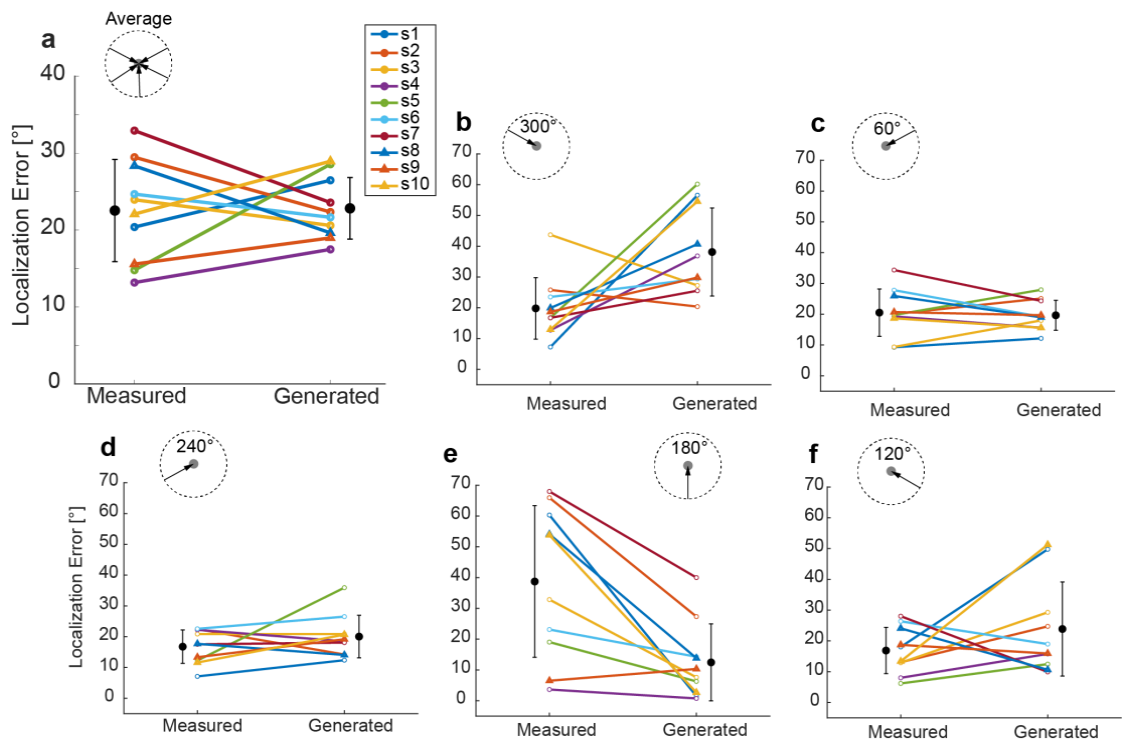

Fig. 5 Localization errors. (a) Averaged localization errors across five directions and trials. Each line represents the errors for individual participants (s1 to S10) when using measured and generated HRTFs. Black error bar represents the mean and standard deviation across participants. (b, c) Errors for the front directions (300° and 60°). (D, E, F) Errors for back directions (240°, 180°, and 120°).

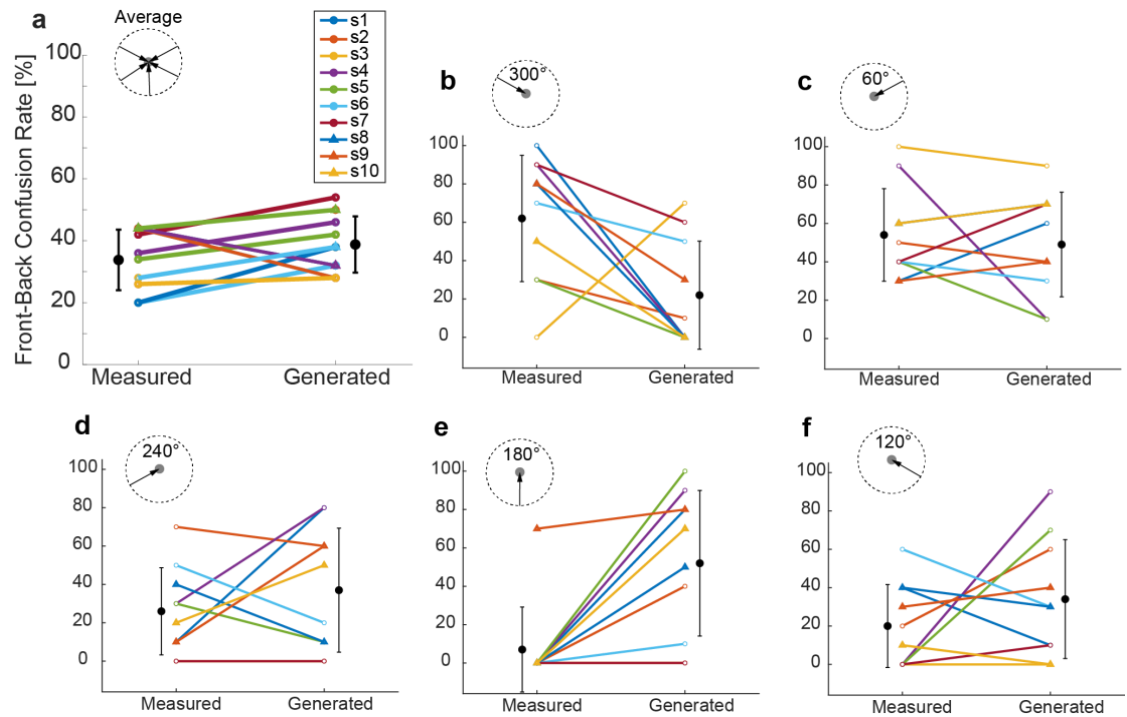

Fig. 6 Front–Back confusion ratio. (a) Averaged front–back confusion ratio across five directions and trials, and the format is similar to the one used in Fig. 5. (b, c) Ratio for front directions (300° and 60°). (d, e, f) Ratio for back directions (240°, 180°, and 120°).

## 4. Discussion

In this study, we proposed a neural network method to generate new HRTFs. The networks were trained on HRTFs from a database to learn the relationship between different directions. As a result, HRTFs for several directions could be generated from a single input HRTF at 0°. This method is expected to reduce the need for actual measurements of HRTFs, making it easier and faster to create personalized HRTFs for virtual auditory displays.

### 4.1 Evaluation of generated HRTFs

In this study, we utilized two metrics, SD and SDR, to evaluate the generated HRTFs, and the cost function was defined as a combination of both metrics. As mentioned previously, SD measures the difference between the generated and measured (ground-truth) HRTFs in the frequency domain. The results of our study indicated that the mean SD for the test of the RIEC database was 6.4 dB, and for the test of the measured dataset, it was 12.6 dB. However, it is unclear how good these SD values are. Another study examined the relationship between SD and subjective evaluation for HRTFs reconstructed by principal component analysis [31], and their results suggest that two HRTFs are subjectively the same when the SD is less than 1 dB, while a difference is perceived when the SD is 5 dB. Based on this evidence, the generated HRTFs using the proposed method may not be a good match, particularly for the new dataset. However, when we conducted the localization test for the generated HRTFs, the accuracy was not significantly worse than that of the measured HRTFs (as shown in Fig. 5 and 6). One possible explanation for this result is that we combined SD and SDR for the cost function. SD only considers the frequency spectrum and does not factor in time domain information. In contrast, SDR measures how closely the generated HRIR matches the measured HRIR in the time domain. Thus, even if the SD is poor, the SDR can be increased so that the generated HRTF is closer to the measured one. Usually, in studies related to HRTF interpolation and extrapolation, the effects of interaural time difference (ITD) and attenuation due to distance in the HRTF are eliminated before SD is evaluated [32]. Therefore, it is challenging to determine the SDR level required for sound localization quantitatively. However, it is well known that ITD, which reflects phase differences, is a critical cue for sound localization, as well as interaural level differences. Hence, considering both SD and SDR is essential for generating accurate HRTFs.

Furthermore, we evaluated the results when using either SD or SDR as the cost function. In the case of the SD cost function, we observed a reduction in SD of 4.7 dB for the test data of the RIEC database and 13.7 dB for the test data of data 2. The extent to which the SD was reduced for the RIEC database was larger when compared to the case where we used both SD and SDR.

However, the result for the newly measured dataset was worse. When we examined SDR, the final SDR was -5.9 dB for the test of data 1 and -4.3 dB for the test of data 2. Thus, the SDR improved without incorporating it into the cost function. Based on these results, the usage of SD as the cost function is likely to overfit the training data, resulting in a degradation of generalization to new dataset.

**4.2 Differences of dataset**

Our results demonstrated a significant difference between the RIEC database and the newly measured dataset. Specifically, the test data for the RIEC database exhibited lower SD and higher SDR, which can be attributed to the fact that the training was conducted using the RIEC database. Despite using the same preprocessing for both datasets, there were differences in the time of the peak responses and the duration of attenuation (as shown in Fig. 5). Additionally, there was a difference of over 9 kHz in the high-frequency ranges of the HRTFs for the newly measured dataset (as shown in Figure 6). According to the database description [10], the data in the RIEC database were measured in a no-reverberation environment. In contrast, the newly measured data was not measured in an environment that suppressed reverberation and reflection as well as the RIEC database. Therefore, the measurement environment likely caused differences in the HRIR/HRTF, particularly in the high-frequency range. We anticipated that the neural networks in our proposed method would learn the relationship between the HRTFs of different directions and be applicable to any HRTF database. However, our results suggest that the proposed method, in its current form, cannot account for environmental or measurement differences. Given that it is desirable to use any HRTF database, this issue should be addressed by eliminating environmental differences in future research.

**4.3 Limitations**

The proposed method still has several limitations, which prevent practical applications. The most important issue is that the SD and SDR in the proposed method do not reach at the level with other interpolation methods. One possible reason is that we directly estimated the HRIR based on the network training using the datasets. The datasets include multiple participants, leading to individual differences. Currently, we have not incorporated a method to address these individual differences. This aspect should be considered in future research. Another issue is the methodological comparison with other techniques. In this study, we examined the results with a cost function of SD only and combined with SD and SDR and showed relative improvement in terms of the temporal profiles when using both SD and SDR. However, it is necessary to compare the results with other methods, including those involving neural networks.

## 5. Conclusion

We proposed a novel method for generating HRTFs in a particular direction from the HRTFs of another direction using temporal convolutional neural networks. Our experiments showed that our method can generate HRTFs for the directions of 60°, 120°, 180°, 240°, and 300° from the 0° HRTF. While the SD and SDR metrics indicated that the generated data still had significant differences from the ground-truth data, the results of the localization test conducted on human participants showed no significant difference between the generated and measured HRTFs. However, further investigation is needed to address the differences observed between different datasets, likely due to variations in measurement environments, and to explore the use of databases or other HRTFs as input to our method for various applications.

**Data availability**

Data will be made available on request.

**Declaration of Competing Interest**

The authors declare no competing interest.

**Acknowledgements**

We would like to thank Editage (www.editage.com) for English language editing.

**Funding**

This research was partially supported by JSPS KAKENHI (19H04112, 21K18304, 22K19809) and Nagai promotion foundation for science of perception.

**Author Contributions**

T.K., I.N., and Y.W. conceived the study and designed the experiment. K.T., and S.Y. developed the experimental setup and K.T. made the experiments and analysed the results with advice from Y.M., I.N., S.Y., and Y.W.. T.K. and I.N. wrote the manuscript. All authors reviewed the final manuscript.